\documentclass[aps,prl,floats,floatfix,twocolumn]{revtex4-1}

\usepackage{graphicx}
\usepackage{epstopdf}
\graphicspath{{./Figs/}{./figs/}}

\usepackage{latexsym}
\usepackage{amsmath}
\usepackage{amssymb}
\usepackage{bm}
\usepackage{wasysym}

\usepackage{color}
\usepackage[colorlinks,linkcolor=blue,urlcolor=blue,bookmarks=true,pdftitle={}]{hyperref}
\usepackage{flafter}

\newcommand{\beq}{\begin{eqnarray}}
\newcommand{\eeq}{\end{eqnarray}} 

\newcommand{\hide}[1]{}
\newcommand{\Eq}[1]{\textcolor{blue}{{Eq.}\!\!~(\ref{#1})}}

\newcommand{\Fig}[1]{\textcolor{blue}{Fig.}\!\!~\ref{#1}}
\definecolor{myred}{rgb}  {0.5,0.0,0.0}
\newcommand{\rmrk}[1]{\textcolor{black}{#1}}

\newcommand{\sect}[1]{{\bf #1.--}}

\newcommand{\Cn}[1]{\begin{center} #1 \end{center}}
\renewcommand{\Re}{\mathrm{Re}}
\renewcommand{\Im}{\mathrm{Im}}

\begin{document}

\title{Adiabatic passage through chaos}
\author{Amit Dey$^1$, Doron Cohen$^2$, Amichay Vardi$^1$} 
\affiliation{
$^1$Department of Chemistry, Ben-Gurion University of the Negev, Beer-Sheva 84105, Israel \\
$^2$Department of Physics, Ben-Gurion University of the Negev, Beer-Sheva 84105, Israel}

%\pacs{
%{71.38.-k, 03.65.Yz, 05.70.Ln, 85.35.Be}}
%03.67.Pp, 75.10.Jm, 87.10.Hk}}
%\date{\today}

\begin{abstract}
We study the process of nonlinear stimulated Raman adiabatic passage within a classical mean-field framework. 
Depending on the sign of interaction, the breakdown of adiabaticity in the interacting non-integrable system, is not related to bifurcations in the energy landscape, but rather to the emergence of quasi-stochastic motion that drains the followed quasi-stationary state. Consequently, faster sweep rate, rather than quasi-static variation of parameters, is better for adiabaticity.
\end{abstract}

\maketitle

%%%%%%%%%%%%%%%%%%%%%%%%%%%%%%%%%%%%%%%%%%%%%%%%%%%%%%%%%%%%%%%%%%%%%%%%%%%%%%%%%%%%%%%%%%%%%%%%%%%%%%%%%%%%%%%%%%%%%%%
\rmrk{The analysis of quasi-static adiabatic processes is a central theme in quantum thermodynamics, coherent control, quantum state engineering, nonlinear and quantum optics,  and nanotechnology. The adiabatic paradigm extends from microscopic systems with few degrees of freedom, through mesoscopic nano-machinary, to macroscopic steam engines. Throughout this vast range of applications, common wisdom has it that "slow is better", i.e. that excitations from the followed adiabatic state can be avoided by slower variation of the system's control parameters. Here, we show that when chaotic stages are encountered during an adiabatic scenario,  slow variation can in fact {\em damage} its efficiency.}

%nonlinear systems
\rmrk{The effect is demonstrated using a minimal example: a stimulated Raman adiabatic passage (STIRAP) \cite{Gaubatz90, Vitanov17} in the presence of interactions. Advances in Bose-Einstein condensation (BEC), nonlinear optics, and  the control of light in coupled waveguides \cite{Fleischer05,Lahini08}, have triggered great interest in the application of adiabatic passage to interacting systems.  The effects of interactions on two-mode adiabatic schemes were studied using various Bose-Hubbard dimer Hamiltonians, \cite{Zobay00,Wu00,Liu02,Liu03,Witthaut06,Witthaut11,Ishkhanyan10,Anglin03,Trimborn10,Altland08,Mannschott09,Chen11,Paredes13,Javanainen99,Yurovsky00, Heinzen00, Ishkanyan04, Altman05, Pazy05, Tikhonenkov06, liu08, liu08b}. The common denominator for all these studies is the quest for energetic stability. The dynamics follows a stationary point (SP) of the instantaneous Hamiltonian $H(x)$, where $x=x(t)$ is a control parameter. This SP, that has some $x$-dependent energy $E[\text{SP}]$, is required to be a local minimum (or a local maximum) of the energy landscape. {\em Nonlinear instability is attributed to the emergence of a separatrix in the energy landscape due to a bifurcation of such a SP}.}

% Three modes
%
\rmrk{The same energetic stability paradigm was adopted for adiabatic passage in the three-mode trimer \cite{Graefe06,Rab08,Bradly12,Polo16,Dupont-Nivet15}:  The \rmrk{SPs} of the energy landscape were found as a function of time, resulting in a bifurcation diagram that reflects topological changes in the energy landscape. Such bifurcations, notably the 'horn' avoided crossing in the nonlinear STIRAP case \cite{Graefe06}, were assumed to cause the breakdown of adiabaticity.   However the three-mode system requires a more careful treatment. While the SPs of systems with more than one degree of freedoms are typically saddle-points of the energy landscape, their {\em dynamical} stability analysis (e.g. via the Bogoliubov formalism \cite{Bogoliubov58,Pethick}) can indicate either stability (real Bogoliubov frequencies) or instability (complex Bogoliubov frequencies). In fact, the full understanding of stability requires a Kolmogorov-Arnold-Moser perspective \cite{Arwas15}. The bifurcation diagram lacks this essential information. For the system under study, Poincare sections are valuable tool for inspecting  the mixed chaotic phasespace structure.}

%%%%%%%%%%%%%%%%%%%%%%%%%%%%%%%%%%%%%%%%%%%%%%%%%%%%%%%%%%%%%%%%%%%%%%%%%%%%%%%%%%%%%%%%%%%%%%%%%%%%%%%%%%%%

\sect{Outline}
We show that the adiabatic passage efficiency is drastically affected by the {\em appearance of chaotic regions, whose existence is not related to the SP bifurcation diagram}. Consequently the analysis of adiabatic passage goes beyond the prevailing energetic stability paradigm. Specifically, reduced efficiency in STIRAP is observed even in the absence of avoided crossings. We establish that the breakdown of adiabaticity occurs during specific time intervals in which the followed-SP becomes immersed in chaotic strips on the {\em same} energy surface. One outcome of this novel breakdown mechanism, is that adiabaticity may be restored by {\em faster} variation of the control parameter, \rmrk{so as to guarantee that the dangerous $x$-interval is traversed before the evolving state has the time to spread along the  chaotic strip.}

%%%%%%%%%%%%%%%%%%%%%%%%%%%%
\begin{figure}[t!]
\centering
\includegraphics[width=3.5in]{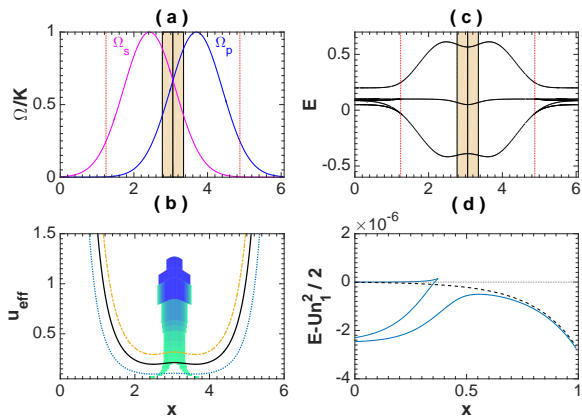} 
\caption{(color online) 
{\bf (a)}~STIRAP pulse scheme. Throughout the manuscript, 
shaded intervals mark the range where chaos leads to breakdown of adiabaticity, 
while vertical dotted lines mark the location of the horn avoided crossings. 
The interaction parameter is ${u=0.2}$ in panels~a,~c.   
Here and in all subsequent figures we set~$\varepsilon{=}0.1$ for the detuning. 
{\bf (b)}~The effective interaction parameter $u_{\rm eff}(x)$
\rmrk{for $u=0.1$ (dotted), $0.2$ (solid), $0.3$ (dash-dotted).} 
\rmrk{The background color indicates the instability of the SP for each $(x,u)$ point: it is white if the Bogoliubov frequencies are real; and colored by green to blue to indicate non-zero magnitude of $\mathrm{Im}(\omega)$, see text.   }
{\bf (c)}~Adiabatic $E[\text{SP}]$ energies. 
The followed state corresponds to the middle curve.  
{\bf (d)}~Emergence of the horn crossing. 
The  $E[\text{SP}]$ of the followed state is zoomed 
for $u=0$ (dotted gray), $u=0.1$ (dashed black), and $u{=}0.101$ (solid blue).
\hfill{} }

\label{f1}
\end{figure}
%%%%%%%%%%%%%%%%%%%%%%%%%%%%

%%%%%%%%%%%%%%%%%%%%%%%%%%%%%%%%%%%%%%%%%%%%%%%%%%%%%%%%%%%%%%%%%%%%%%%%%%%%%%%%%%%%%%%%%%%%%%%%%%%%%%%%%%%%
%%%%%%%%%%%%%%%%%%%%%%%%%%%%%%%%%%%%%%%%%%%%%%%%%%%%%%%%%%%%%%%%%%%%%%%%%%%%%%%%%%%%%%%%%%%%%%%%%%%%%%%%%%%%
\sect{STIRAP}
Many-body STIRAP is modelled by the time-dependent Bose-Hubbard trimer Hamiltonian \cite{Graefe06,Eilbeck95,Hennig95,Franzosi03,Flach97,Nemoto00,Franzosi02,Hiller06,Tikhonenkov13} for $N$ particles in three second-quantized modes:
\beq \label{ham}
%\mathcal{H} \ =&& \ \mathcal{E}\hat{n}_2 \ + \ \frac{U}{2}\sum_{j=1}^3\hat{n}^2_i 
\mathcal{H} \ =&& \ {\cal E}\hat{n}_2 \ + \ \frac{U}{2}\sum_{j=1}^3\hat{n}^2_i 
\\ \nonumber
&-& \frac{1}{2}\left(\Omega_p(x)\hat{a}_2^{\dagger}\hat{a}_1 + \Omega_s(x)\hat{a}_3^{\dagger}\hat{a}_2+h.c.\right)~.
\eeq
%
%
%\beq \label{ham}
%&& \mathcal{H} \ = \ \mathcal{E}\hat{n}_2 \ + \ \frac{U}{2}\sum_{j=1}^3\hat{n}^2_i \ + \ W(x(t)) 
%\\ \nonumber
%&& \ \ W(x) = -\frac{K}{2}\left(\Omega_p(x)\hat{a}_2^{\dagger}\hat{a}_1 + \Omega_s(x)\hat{a}_3^{\dagger}\hat{a}_2+h.c.\right)
%\eeq
%
Here,  $\hat{a}_j$, $\hat{a}_j^\dag$ are bosonic operators with associated occupations $\hat{n}_j\equiv \hat{a}_j^{\dagger}\hat{a}_j$.  The interaction strength is~$U$, while ${\cal E}$ is equivalent to the one-photon detuning of the optical scheme \cite{Gaubatz90, Vitanov17}. 
In STIRAP, the couplings are Gaussian Stokes and Pump pulses $\Omega_{s,p}(x)=Ke^{-\left(x-x_{s,p}\right)^2}$ which depend on the dimensionless parameter~$x$.
The standard realization is a simple constant-rate sweep $x(t) = t/\tau$, 
with a `counterintuitive' sequence \rmrk{${ x_p-x_s > 0 }$, as shown in \Fig{f1}a.} 
The system is prepared in the first mode $(n_1(0)=N)$. 
For $U=0$, an adiabatic sweep transfers the population 
to the third mode ($n_3(\infty)=N$) by following 
a coherent dark eigenstate that does not project
on the intermediate mode at any time ($n_2(t)=0$). 
The studied effect is the breakdown of this adiabatic 100\% efficiency 
in the presence of repulsive interactions ($U>0$).

%%%%%%%%%%%%%%%%%%%%%%%%%%%%%%%%%%%%%%%%%%%%%%%%%%%%%%%%%%%%%%%%%%%%%%%%%%%%%%%%%%%%%%%%%%%%%%%%%%%%%%%%%%%%
%%%%%%%%%%%%%%%%%%%%%%%%%%%%%%%%%%%%%%%%%%%%%%%%%%%%%%%%%%%%%%%%%%%%%%%%%%%%%%%%%%%%%%%%%%%%%%%%%%%%%%%%%%%%
\sect{Classical dynamics}
In classical mean-field theory, field operators $\hat{a}_j$ are replaced 
by c-numbers  $a_j \equiv \sqrt{n_j} e^{i \phi_j}$.   
Rescaling $a_j \mapsto a_j/\sqrt{N}$, and  $t\mapsto Kt$,  
\rmrk{and defining $P_j=|a_j|^2$, 
we obtain the nonlinear Schr\"odinger equations \cite{Graefe06} 
${i\dot{\mathbf a}=\left({\cal H}_{0} + u{\mathcal P}\right){\mathbf a}}$, 
where 
\beq
&& \mathcal{H}_{0}{=}
\begin{pmatrix}
0&-\kappa_p/2&0\\
-\kappa_p/2&\varepsilon&-\kappa_s/2\\
0&-\kappa_s/2&0
\end{pmatrix},~
\mathcal{P}{=}
\begin{pmatrix}
P_1&0&0\\
0&P_2&0\\
0&0&P_3
\end{pmatrix}. \ \ \ \ 
\eeq
The dimensionless parameters} are the interaction ${u=NU/K}$,  
the detuning ${\varepsilon={\cal E}/K}$, 
and the couplings ${\kappa_{p,s}=\Omega_{p,s}/K}$.  
We also define the effective nonlinearity ${u_{\rm eff}(x)=u/(\kappa_p^2(x)+\kappa_s^2(x))^{1/2}}$. 
The latter is largest at the beginning and at the end of the sweep, where the linear coupling terms are small, see \Fig{f1}b.

%%%%%%%%%%%%%%%%%%%%%%%%%%%%%%%%%%%
\begin{figure}[t!]
\centering
\includegraphics[width=3.5in]{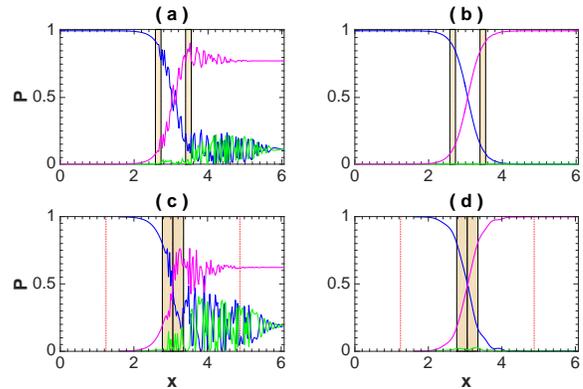} 
\caption{(color online) 
Evolution of the site populations versus~$x(t)$.
Locations of the horn crossings (if exist) 
and of the chaotic intervals are marked.      
{\bf (a)}~Failure of STIRAP in the absence of SP bifurcations: 
here $u=0.8\varepsilon$ is {\em below} the critical value for obtaining the horn crossing.
The sweep rate is $\dot{x}/K=6\times10^{-5}$.  
{\bf (b)}~Recovery of adiabatic passage with {\em increased} sweep rate 
($\dot{x}/K=6\times10^{-4}$) during chaotic intervals.
{\bf (c)}~~Failure of STIRAP for $u=2\varepsilon$, 
with initial conditions that bypasses the horn crossing:
the process is launched at the adiabatic state {\em after} the avoided crossing. 
Sweep rate is $\dot{x}/K=6\times10^{-4}$. 
{\bf (d)}~For same $u$, efficiency 
is recovered due to faster sweep ($\dot{x}/K=4\times10^{-2}$). 
\hfill{} }
\label{f2}
\end{figure}
%%%%%%%%%%%%%%%%%%%%%%%%%%%%%%%%%%%%

%%%%%%%%%%%%%%%%%%%%%%%%%%%%%%%%%%%%
\sect{Bifurcation diagram}
\rmrk{The steady states of our system (at fixed~$x$) are the SPs of the grand canonical Hamiltonian ${\mathcal{H}-\mu N}$, satisfying $i\dot{\bf a}=\mu{\bf a}$, where $\mu$ is identified as the chemical potential.
The solution of this equation has been presented in~\cite{Graefe06}.
The adiabatic energies ${E[\text{SP}]}$ are the value of ${\mathcal{H}}$ at the SPs.}
For ${u=0}$ there are three SPs, corresponding to the adiabatic eigenstates of  linear STIRAP \cite{Vitanov17}.
In the presence of interaction the SPs bifurcate if the effective interaction $u_{\text{eff}}(x)$ is large enough, 
i.e. at early and late times, as shown in \Fig{f1}c.   
For ${u>\varepsilon}$ the 'horn' avoided crossing appears \cite{Graefe06}, see \Fig{f1}d. 
As $u$ increases, more SPs emerge.

Careful inspection shows that the nonlinear breakdown of adiabaticity goes beyond the bifurcation diagram analysis. In \Fig{f2}a, inefficient transfer at low $\dot x$ is obtained even for ${u<\varepsilon}$, where no horn crossing is present. Population oscillations, indicating non-adiabaticity, are boosted only during the marked intervals in \Fig{f2}a, for which the adiabatic bifuraction diagram exhibits no special nonlinear features. 
Moreover, as shown in \Fig{f2}c, while for ${u>\varepsilon}$ the horn crossing does appear in an early stage, adiabaticity breaks down even if the system is initiated {\em after} it. Here too, the growth of population oscillations does not correlate with the avoided crossing or any other feature in the bifurcation diagram. 

Another unique finding is the dependence of transfer efficiency on sweep rate. Oddly, the efficiency increases for {\em faster} sweep rates. In fact, as demonstrated in \Fig{f2}b, 
adiabaticity can be restored by speeding up the sweep process only during the marked intervals mentioned above. This prescription obviously has nothing to do with bifurcations of stationary solutions.

%%%%%%%%%%%%%%%%%%%%%%%%%%%%%%%%%%%%
\begin{figure}[t!]
\centering
\includegraphics[width=\hsize]{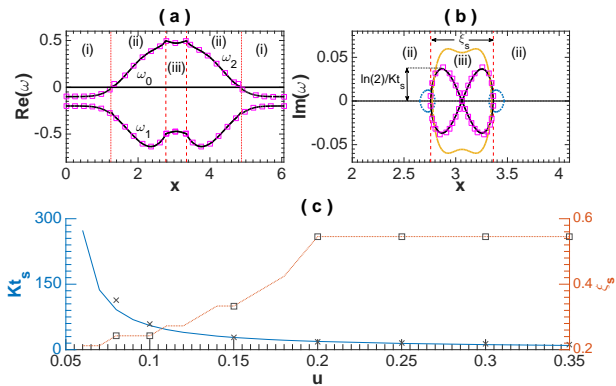} 

\caption{(color online)
\rmrk{Stability analysis: 
{\bf (a)} $\Re(\omega_j$) for $u{=}0.2$. The followed SP is swapped from an energy maximum in region~(i) to a saddle point in regions~(ii) and~(iii). 
{\bf (b)} $\Im(\omega_j)$  for $u=0.1$ (dotted), $0.2$ (solid), $0.3$ (dash-dotted).
It becomes non-zero in region~(iii), implying loss of dynamical stability.
Square markers in (a),(b) denote approximate analytical solutions \cite{sup} for $u{=}0.2$.  
The parameters $\xi_s$ and $t_s$ are extracted as shown. 
{\bf (c)} Dependence of $\xi_s$ (dotted) and $t_s$ (solid) on $u$. 
Markers denote $\xi_s$ ($\square$)  and $t_s$ ($\times$) as obtained 
from spreading simulations of a semiclassical cloud \cite{sup} .}
\hfill{} }

\label{f3}
\end{figure}
%%%%%%%%%%%%%%%%%%%%%%%%%%%%%%%%%%%%

%%%%%%%%%%%%%%%%%%%%%%%%%%%%%%%%%%%%%%%%%%%%%%%%%%%%%%%%%%%%%%%%%%%%%%%%%%%%%%%%%%%%%%%%%%%%%%%%%%%%%%%%%%%%
%%%%%%%%%%%%%%%%%%%%%%%%%%%%%%%%%%%%%%%%%%%%%%%%%%%%%%%%%%%%%%%%%%%%%%%%%%%%%%%%%%%%%%%%%%%%%%%%%%%%%%%%%%%%
\sect{Stability Analysis}
%
%\rmrk{Dynamical stability is evaluated from Bogoliubov stability analysis around the followed SP. The magnitude of the imaginary part of the resulting characteristic frequencies is plotted in Fig.~%\ref{f1}b. The intervals in which a faster sweep rate gives better efficiency correspond precisely to the regions of dynamical instability.}
%
\rmrk{The followed SP's stability is determined by solving the Bogoliubov equations \cite{Bogoliubov58,Pethick} for its quasiparticle modes $({\mathbf u}_j, {\mathbf v}_j)$ and frequencies $\omega_j$. 
Defining ${\cal L}={\cal H}_0+2u{\mathcal{P}(\mathbf{a}_\mathrm{SP})}-\mu/K$, and ${\cal M}=-u\mathcal{P}(\mathbf{a}_\mathrm{SP})$, 
where $\mathbf{a}_\mathrm{SP}$ is the state vector at the stationary point, these equations read: 
\beq\label{bog}
\label{bog}
\begin{pmatrix}
{\cal L}&{\cal M}\\
-{\cal M}&-{\cal L}
\end{pmatrix}
\begin{pmatrix}
{\mathbf u}_j\\
{\mathbf v}_j
\end{pmatrix}
=
\omega_j
\begin{pmatrix}
{\mathbf u}_j\\
{\mathbf v}_j
\end{pmatrix}.
\eeq
Energetic stability is determined by the signs of~$\Re(\omega_j)$, while dynamical instability is indicated by non-vanishing $\Im(\omega_j)$. An analytical approximation for the Bogoliubov frequencies \cite{sup} is in excellent agreement with direct numerical diagonalization, see \Fig{f3}. The resulting frequencies include the zero mode $\omega_0{=}0$ due to global gauge symmetry \cite{Bogoliubov58,Pethick}. From the remaining frequencies $\omega_{1,2}$ it is clear that while the horn crossing amounts to a transition from a self-trapped energy maximum ($\omega_{1,2}{<}0$) to a saddle point ($\omega_1{<}0$, $\omega_2{>}0$), dynamical instability only appears later,  in precise agreement with the ``adiabaticity killing grounds" of \Fig{f2}. The breakdown of STIRAP efficiency is thus not due to {\em energetic instability}, but rather due to {\em dynamical instability}.  From these plots, we find the width of the unstable region $\xi_s$, and the characteristic instability time $t_s=\ln{2}/\max(\omega)$ at which the fluctuations are doubled. These parameters agree well with numerical simulations of the spreading of a semiclassical cloud around the SP \cite{sup}.}

%%%%%%%%%%%%%%%%%%%%%%%%%%%%%%%%%%%%
\begin{figure}[t!]
\centering
\includegraphics[width=\hsize]{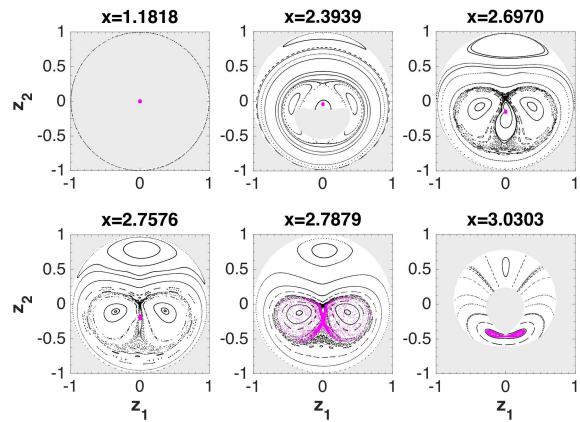} 

\caption{(color online) 
Poincare sections for the fixed $x$ Hamiltonian at representative values of~$x$. Here $u{=}0.22$. The energy in all panels is $E=E[\text{SP}]$ of the followed-SP. The cross section is taken through the ${n_2=n_2[\text{SP}]}$ plane of the 3D energy surface. 
We use polar coordinates \rmrk{$z_1=r\cos\varphi$, $z_2=r\sin\varphi$}, 
where ${r=(1-(n/N))/2}$. 
Magenta dots correspond to a semiclassical cloud, initially localized around the followed-SP. Gray shading marks energetically forbidden regions. Note that these panels depict the adiabatic sequence up to the middle point ${x\sim3}$. The Poincare sections at later times mirror the presented panels, and contain a second chaotic interval. 
\hfill{} }

\label{f4}
\end{figure}
%%%%%%%%%%%%%%%%%%%%%%%%%%%%%%%%%%%%

%%%%%%%%%%%%%%%%%%%%%%%%%%%%%%%%%%%%%%%%%%%%%%%%%%%%%%%%%%%%%%%%%%%%%%%%%%%%%%%%%%%%%%%%%%%%%%%%%%%%%%%%%%%%
%%%%%%%%%%%%%%%%%%%%%%%%%%%%%%%%%%%%%%%%%%%%%%%%%%%%%%%%%%%%%%%%%%%%%%%%%%%%%%%%%%%%%%%%%%%%%%%%%%%%%%%%%%%%
\sect{Passage through chaos}
In \Fig{f4}, we show representative Poincare sections for several~$x$ values during the adiabatic passage. \rmrk{The Bose-Hubbard trimer is a two freedom system (two population imbalances and two relative phases serving as conjugate coordinates), hence its phase space is 4D and the fixed energy surfaces are 3D. For a given $N$ and $E$ our dynamical coordinates are the middle site occupation $n_2$, the population imbalance ${n=n_1{-}n_3}$, and the relative phase ${\varphi=\varphi_1{-}\varphi_3}$. All the trajectories belong to the pertinent energy surface $E=E[\text{SP}]$. A trajectory is sampled each time that it intersects the plane ${n_2=n_2[\text{SP}]}$. Accordingly we get a section whose coordinates are ${\mathbf{z}=(\varphi,n)}$. These are displayed as polar coordinates in \Fig{f4}.}
Note that the observed structures do not reflect the topography of the energy landscape, but correspond to various periodic orbits, invariant tori, and chaotic regions {\em on the same energy surface}. The plotted sections contain a single SP that supports the followed adiabatic eigenstate, while the other 'fixed-points' are in fact periodic orbits. In each section, we plot the evolution of a cloud that is launched around the followed-SP.

The sequence of Poincare sections reveals the \rmrk{source of dynamical instability}. At early times ($x=1.1818$) the dynamics is interaction-dominated and the evolution is restricted to self-trapped trajectories. Appropriately for an energy maximum, the followed SP is surrounded by an energetically forbidden region (gray). After the horn crossing the followed SP is an energy saddle, the forbidden region disappears, and an intermediate non-linear resonance shows up as a `belt' in the Poincare section ($x=2.3939$). 
At larger~$x$, the belt expands, and a chaotic strip is formed along its border  ($x=2.6970$). The enclosed 'island', containing the followed-SP, shrinks down until the SP hits the chaotic strip ($x=2.7576$). The dynamical instability intervals correspond to the embedding of the followed-SP in the chaotic strip, resulting in the quasi-stochastic spreading of the initially localized distribution over the chaotic region ($x=2.7879$). 
The entire progression takes place on a single 3D energy surface, and has no trace in the adiabatic energy diagram. 

%%%%%%%%%%%%%%%%%%%%%%%%%%%%%%%%%%%%%%%%%%%%%%%%%%%%%%%%%%%%%%%%%%%%%%%%%%%%%%%%%%%%%%%%%%%%%%%%%%%%%%%%%%%%
%%%%%%%%%%%%%%%%%%%%%%%%%%%%%%%%%%%%%%%%%%%%%%%%%%%%%%%%%%%%%%%%%%%%%%%%%%%%%%%%%%%%%%%%%%%%%%%%%%%%%%%%%%%%
\sect{Adiabaticity threshold}
%
%The {\em passage through chaos mechanism} also explains why the speed-up of the sweep during the chaotic intervals can restore the STIRAP efficiency. 
The draining of the SP region can be avoided if the chaotic interval $\xi_s$ is traversed on a shorter time scale than the instability time $t_s$. 
Thus, a {\em low sweep rate} adiabaticity threshold should exist. Combining with the standard adiabaticity condition we deduce that high STIRAP efficiency is maintained for   
\beq
\frac{\xi_s}{t_s} \ \  < \ \ \dot{x} \ \ < \ \  \frac{1}{3\pi} K~.
\label{apc}
\eeq
The upper limit is required for 96\% efficiency \cite{Vitanov17} and ensures small probability for non-adiabatic transitions in the transverse (energy) direction. If $\dot{x}$ is constant throughout the evolution, the adiabaticity threshold condition translates into ${\tau < t_s/x_s}$ for the sweep time. For larger $u$, the $\xi_s$ range becomes larger, while $t_s$ becomes smaller (see \Fig{f3}c). Consequently the adiabaticity threshold is  monotonically increasing as a function of~$u$.

%%%%%%%%%%%%%%%%%%%%%%%%%%%%%%%%%%%%
\begin{figure}[t!]
\centering
\includegraphics[width=3.5in]{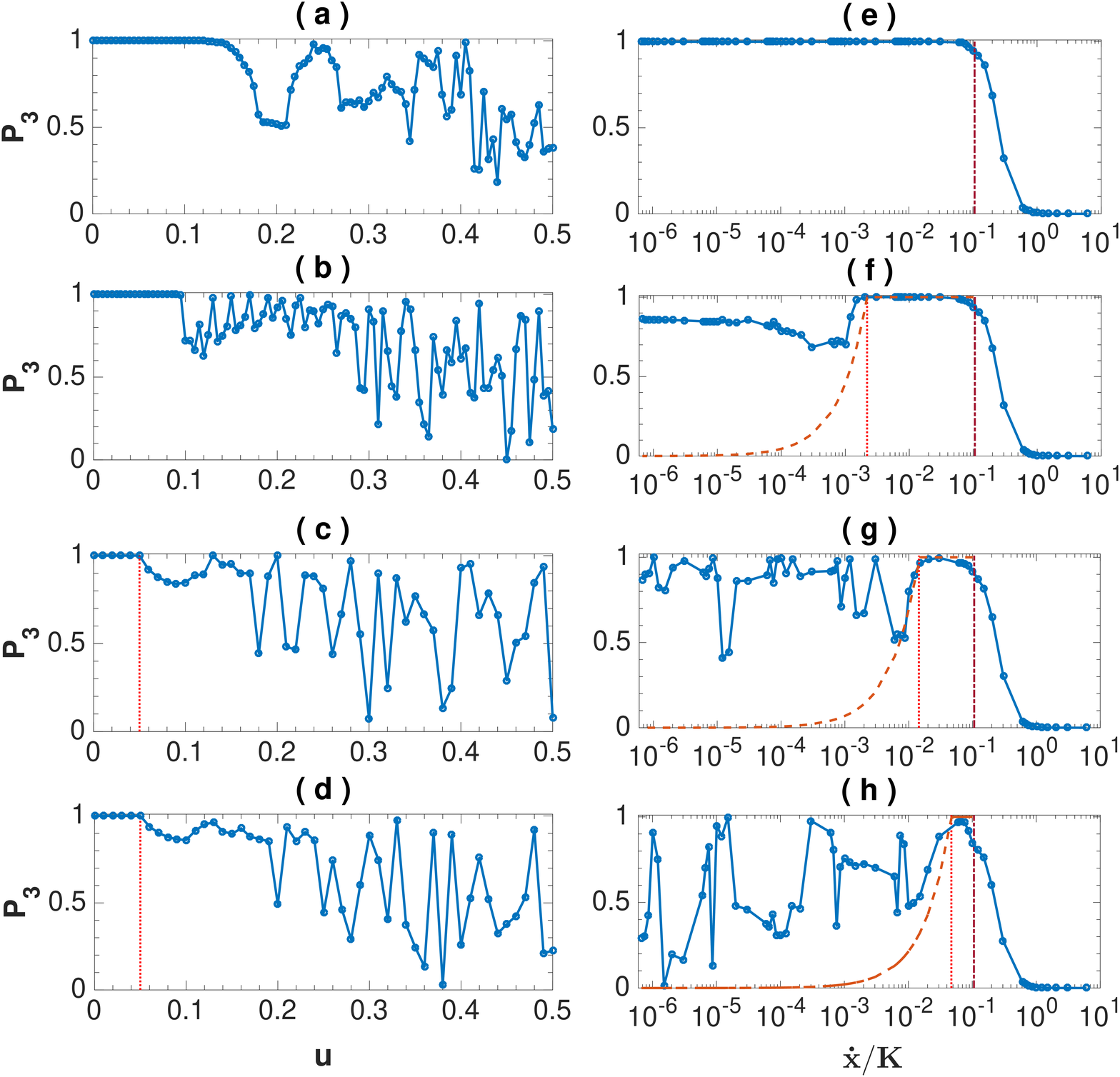}

\caption{(color online)  
STIRAP efficiency. $P_3$ is the population fraction in the target state at the end of the nonlinear STIRAP. Panels a-d show $P_3$ as a function of~$u$ for $\dot{x}/K=6{\times}10^{-3}, 6{\times}10^{-4}, 6{\times}10^{-6}, 6{\times}10^{-7}$, respectively. 
Vertical lines in c,d mark the chaoticity threshold ${u=\varepsilon/2}$. 
Panels e-h present $P_3$ as a function of the sweep rate $\dot{x}$ for $u=0.05,0.1,0.2,0.3$, respectively. Vertical lines mark the range of adiabaticity (condition \Eq{apc}). The estimated SP survival probability \Eq{SPS} is plotted as orange dashed line.
\hfill{} }

\label{f5}
\end{figure}
%%%%%%%%%%%%%%%%%%%%%%%%%%%%%%%%%%%%

%%%%%%%%%%%%%%%%%%%%%%%%%%%%%%%%%%%%%%%%%%%%%%%%%%%%%%%%%%%%%%%%%%%%%%%%%%%%%%%%%%%%%%%%%%%%%%%%%%%%%%%%%%%%
%%%%%%%%%%%%%%%%%%%%%%%%%%%%%%%%%%%%%%%%%%%%%%%%%%%%%%%%%%%%%%%%%%%%%%%%%%%%%%%%%%%%%%%%%%%%%%%%%%%%%%%%%%%%
\sect{Horn vs Belt resonance} 
The horn avoided crossing~\cite{Graefe06} is a $1{:}1$ resonance \rmrk{(frequency of first site matches the frequency of the second due to interaction)}. 
It is born provided ${u>\varepsilon}$ such that the condition ${U n_1 = {\cal E}}$ can be satisfied. 
We realize that there is also a nonlinear $2{:}1$ resonance that manifests itself if ${u>\varepsilon/2}$,  
\rmrk{(frequency of the first site is half the detuning)}. It shows up
in the Poincare section as a belt that consist of two islands.   
This belt is born far away from the followed SP, but nevertheless it can choke the SP in a later stage (Fig.~\ref{f4}).      

We note that weak non-adiabatic effects due to horn resonance can be detected as well, but for ${u>0}$, as discussed above, they are overwhelmed by the passage-through-chaos mechanism. In contrast, Poincare sections for ${u<0}$ (not presented) show that the SP does not go through the chaotic strip of the non-linear belt. Consequently, in the latter case, the passage-through-chaos mechanism becomes irrelevant, and the failure of STIRAP is purely due to the horn crossing.

%%%%%%%%%%%%%%%%%%%%%%%%%%%%%%%%%%%%%%%%%%%%%%%%%%%%%%%%%%%%%%%%%%%%%%%%%%%%%%%%%%%%%%%%%%%%%%%%%%%%%%%%%%%%
%%%%%%%%%%%%%%%%%%%%%%%%%%%%%%%%%%%%%%%%%%%%%%%%%%%%%%%%%%%%%%%%%%%%%%%%%%%%%%%%%%%%%%%%%%%%%%%%%%%%%%%%%%%%
\sect{STIRAP efficiency}
In \Fig{f5}, we plot the STIRAP efficiency as a function of $u$ for several values of $\dot{x}$, as well as the complimentary dependence on~$\dot{x}$ at fixed~$u$. The shrinking of the high efficiency $u$~range  as $\dot{x}$ is decreased, reflects the breakdown of adiabaticity due to the passage-through-chaos mechanism. In the adiabatic regime (panels~c-d), the range of ${\sim}100\%$ efficiency is restricted by the {\em chaoticity threshold} (${|u|<|\varepsilon|/2}$) below which no stochastic strips are formed. We note that a similar plot in Ref.~\cite{Graefe06} corresponds to an intermediate value of~$\dot{x}$, hence it does not represent the adiabatic regime. 

Looking at the right panels of \Fig{f5}, we see that below the chaoticity limit (panel~e) there is no breakdown in the slow sweep limit, and the efficiency is monotonically decreasing with the rate, just as in the linear case. Once chaos sets in (panels~f-h), high efficiency can still be maintained if condition \Eq{apc} is satisfied. As~$u$ is further  increased, the high efficiency range between the slow and fast sweep boundaries shrinks, until the two inequalities of \Eq{apc} can not be satsified simultaneously (panel~h).

The transfer probability $P_3$ can be written as a sum ${ P_{\text{surv}}+P_{\text{scat}} }$, 
where $P_{\text{surv}}$ is the probability for survival in the SP region, 
while $P_{\text{scat}}$ the scattered component. The former can be estimated 
as follows: The spreading trajectories in the stochastic region 
have frequencies ${\omega \in  [0,1/t_s]}$ with roughly uniform distribution. 
Trajectories that survive in the SP region satisfy ${ \omega \times (\xi_s/\dot{x}) < 1 }$, 
hence their fraction is  
\beq \label{SPS}
 P_{\text{surv}} \ \ = \ \ \min\{ (t_s/\xi_s)\dot{x}, \ 1  \}
\eeq
This estimate can serve as a lower bound for the STIRAP efficiency as illustrated in \Fig{f5} panels~f-h.

%%%%%%%%%%%%%%%%%%%%%%%%%%%%%%%%%%%%%%%%%%%%%%%%%%%%%%%%%%%%%%%%%%%%%%%%%%%%%%%%%%%%%%%%%%%%%%%%%%%%%%%%%%%%
%%%%%%%%%%%%%%%%%%%%%%%%%%%%%%%%%%%%%%%%%%%%%%%%%%%%%%%%%%%%%%%%%%%%%%%%%%%%%%%%%%%%%%%%%%%%%%%%%%%%%%%%%%%%
\sect{Conclusions}
The physics of three-mode adiabatic passage schemes is more intricate than that of the nonlinear Landau-Zener paradigm. The latter relies entirely  on {\em energetic stability}, which is endangered by bifurcations of the followed-SP. By contrast, the failure of adiabatic passage in non-integrable systems is related to {\em dynamical instability} on a single multi-dimensional energy surface, containing both quasi-integrable and chaotic regions. Consequently, adiabatic-passage efficiency can be improved by faster variation of the control parameters. 
\rmrk{The role of chaos as an optional tool to control the outcome of a STIRAP scheme has been pointed in \cite{Na04,Na05} in a different context: there the chaos was due to the laser frequencies, and the analysis was based on Floquet theory that goes beyond the traditional rotating-wave approximation of \Eq{ham}.}

%%%%%%%%%%%%%%%%%%%%%%%%%%%%%%%%%%%%%%%%%%%%%%%%%%%%%%%%%%%%%%%%%%%%%%%%%%%%%%%%%%%%%%%%%%%%%%%%%%%
%%%%%%%%%%%%%%%%%%%%%%%%%%%%%%%%%%%%%%%%%%%%%%%%%%%%%%%%%%%%%%%%%%%%%%%%%%%%%%%%%%%%%%%%%%%%%%%%%%%

\clearpage

\sect{Acknowledgements}
This research was supported by the Israel Science Foundation (Grant  No. 283/18)

%%%%%%%%%%%%%%%%%%%%%%%%%%%%%%%%%%%%%%%%%%%%%%%%%%%%%%%%%%%%%%%%%%%%%%%%%%%%%%%%%%%%%%%%%%%%%%%%%%%
%%%%%%%%%%%%%%%%%%%%%%%%%%%%%%%%%%%%%%%%%%%%%%%%%%%%%%%%%%%%%%%%%%%%%%%%%%%%%%%%%%%%%%%%%%%%%%%%%%%

%\end{document}

%%%%%%%%%%%%%%%%%%%%%%%%%%%%%%%%%%%%%%%%%%%%%%%%%%%%%%%%%%%%%%%%%%%%%%%%%%%%%%%%%%%%%%%%%%%%%%%%%%%%%%%%%%%
%%%%%%%%%%%%%%%%%%%%%%%%%%%%%%%%%%%%%%%%%%%%%%%%%%%%%%%%%%%%%%%%%%%%%%%%%%%%%%%%%%%%%%%%%%%%%%%%%%%%%%%%%%%

\clearpage

\onecolumngrid

\pagestyle{empty}
\renewcommand{\thefigure}{S\arabic{figure}}
\setcounter{figure}{0}

\Cn{
{\Large \bf Adiabatic passage through chaos} \\
Amit Dey, Doron Cohen, Amichay Vardi \\
{SUPPLEMENTARY MATERIAL} \\ \
}

%%%%%%%%%%%%%%%%%%%%%%%%%%%%
\section{Chaos vs. horn}

\begin{figure} %  figure placement: here, top, bottom, or page
   \centering
   \includegraphics[width=12cm]{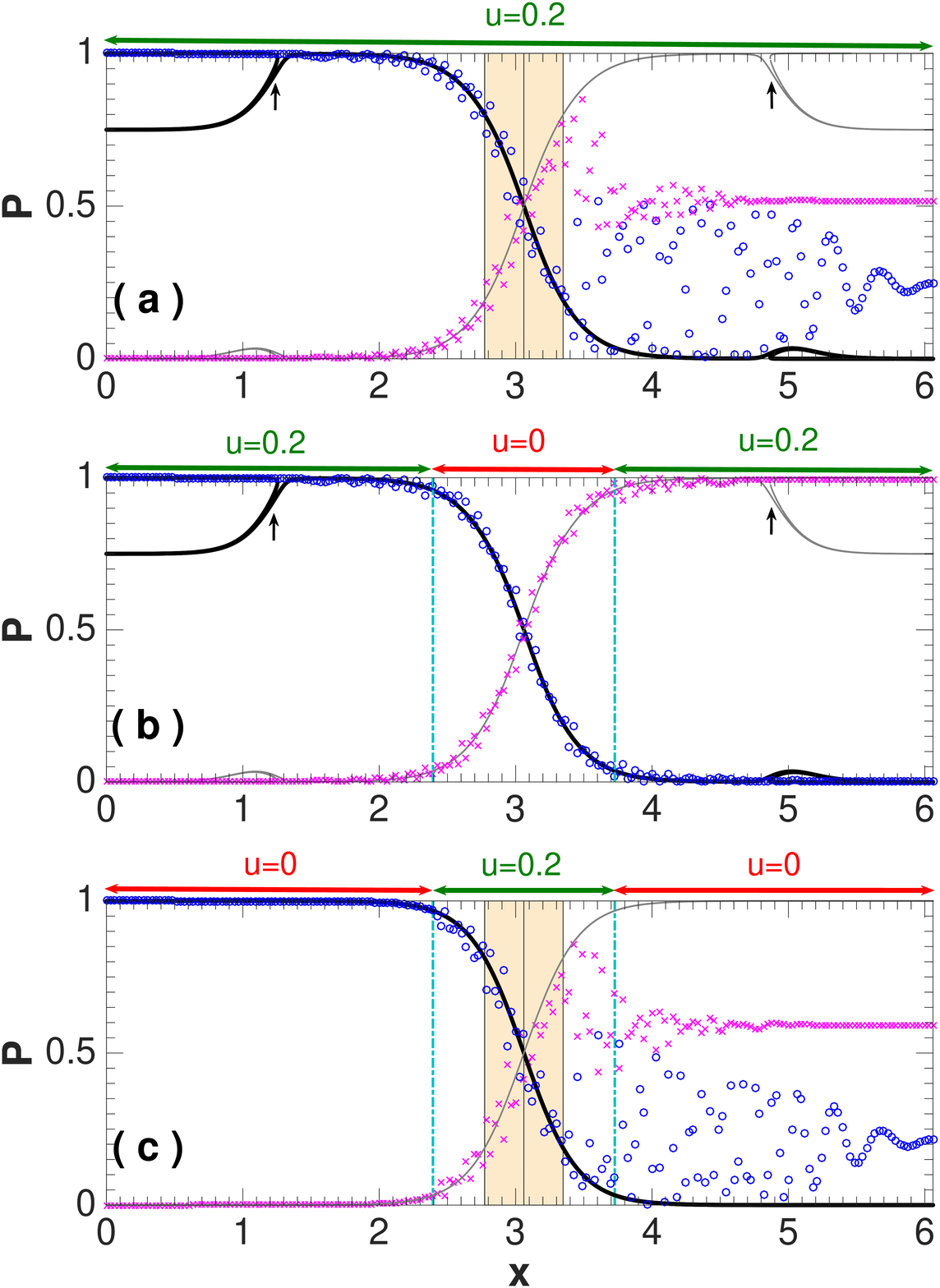} 
   \caption{Population dynamics for $\dot{x}=6\times10^{-3}$ and $u=0.2$. {\bf (a)} {\em Horn and chaos}: The interaction is  on throughout the process. {\bf (b)} {\em Horn without chaos}: The interaction is turned off between the dash-dotted vertical line so the horn crossings exist but there is no chaotic interval. {\bf (c)} {\em Chaos without horn}: The interaction is only on between the dash-dotted vertical lines so there is no avoided crossing but a chaotic interval exists. Thick black and thin grey lines correspond respectively, to the projections of the followed adiabatic eigenstate onto the initial and target sites. Arrows and shaded areas mark respectively the horn crossings and the chaotic intervals when either or both exist.}
   \label{fig1_s}
\end{figure}

The breakdown of adiabaticity during nonlinear STIRAP is depicted In \Fig{fig1_s}a. This is the same as Fig.~2c of the manuscript, only the system is launched before the horn avoided crossing so it is not bypassed.  Comparing the population dynamics with the site-projections of the followed SP, we see that the horn crossing itself is easily traversed and adiabaticity only breaks down later during the chaotic interval. To sharpen the distinction between the horn crossing and the chaotic interval as the cause of  the breakdown, we present two complimentary scenarios for the same sweep rate: In \Fig{fig1_s}b we turn off the interaction during the time at which chaotic interval would have existed. Since the interaction is on during the horn crossings, this scenario corresponds to a 'pure horn effect', with the result that the system follows the adiabatic eigenstate. By contrast, in \Fig{fig1_s}c, the interaction is only turned on during the chaotic interval so that the horn crossings are eliminated.  The breakdown of adiabaticity in this 'pure chaos' scenario, along with the adiabaticity of the 'pure horn' scenario of \Fig{fig1_s}b, constitute an unequivocal demonstration that the cause is the chaotic interval rather than the horn crossing. 

\begin{figure} %  figure placement: here, top, bottom, or page
   \centering
   \includegraphics[width=7cm]{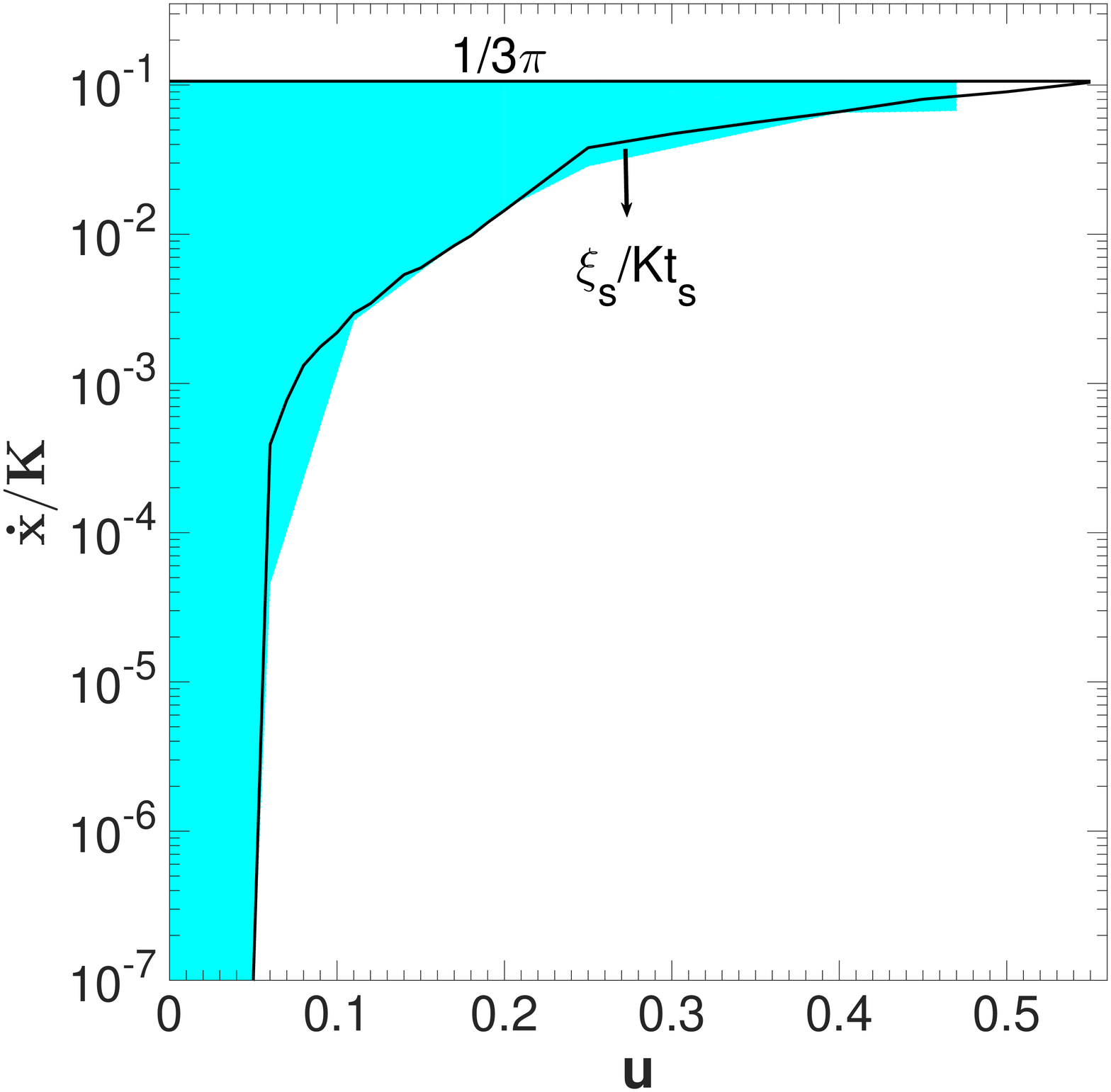} 
   \caption{Adiabaticity diagram. The shaded region corresponds to the parameter range where $P_3>96\%$ at the end of the numerical STIRAP simulations.  Solid lines mark the theoretical boundaries of Eq.~(4). The lower boundary is based on the parameters $\xi_s,t_s$ of the dynamical instability.}
   \label{fig2_s}
\end{figure}

This observation holds throughout the parameter range, as long as the interaction is repulsive. In \Fig{fig2_s}, we plot the range of high (adiabatic) transfer efficiency throughout the $(u, {\dot x})$ parameter space. It is clear that the adiabatic range is determined by the condition of Eq.~(4) in the manuscript. This condition is based purely on the characteristic parameters of the chaotic interval and has no relation to the horn crossing. Therefore we conclude that the horn crossing does not play a role in the breakdown of adiabaticity when $u>0$. In the following item we explain why this is the case.

%%%%%%%%%%%%%%%%%%%%%%%%%%%%

%%%%%%%%%%%%%%%%%%%%%%%%%%%%

%%%%%%%%%%%%%%%%%%%%%%%%%%%%
\section{The Horn crossing in Phase Space}
\begin{figure} %  figure placement: here, top, bottom, or page
   \centering
   \includegraphics[width=10cm]{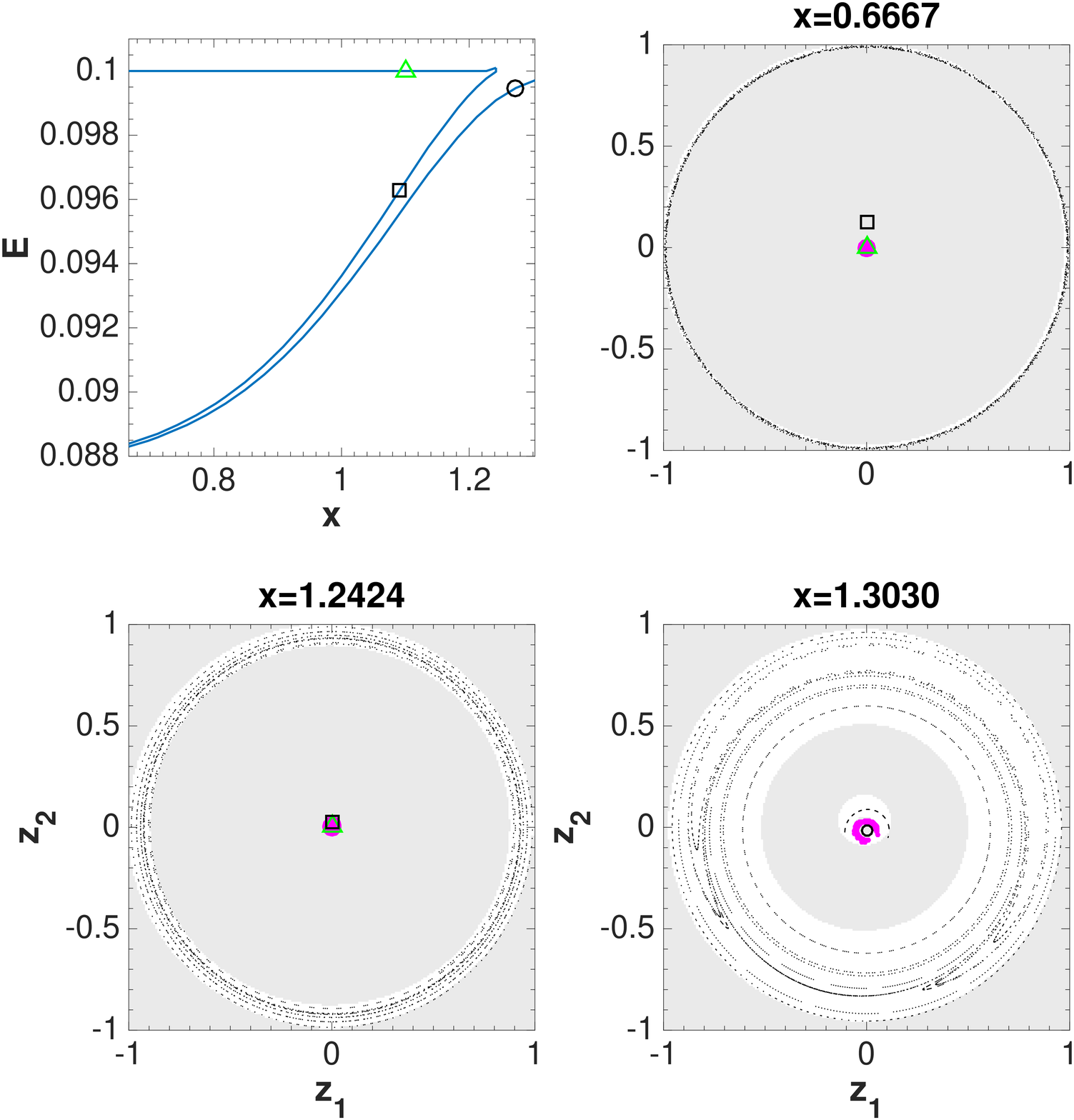} 
   \caption{Phase space dynamics of the horn crossing. The participating SPs are marked in the top left panel.  At the horn crossing, the followed SP changes from the energy maximum $\triangle$ to the saddle point $\circ$. The remaining panels show the evolution of a localized semiclasical cloud, launched around the $\triangle$ SP before the crossing. All Poincare sections are plotted at the cloud energy as it evolves. Before the crossing (top right), the followed SP ($\triangle$) is an energy maximum  surrounded by a forbidden region that contains the lower energy SPs ($\square,\circ$). This maximum is annihilated at the horn crossing as it merges with the $\square$ SP (bottom left). The close proximity to the saddle point $\circ$, allows a smooth $\triangle\rightarrow\circ$ swap  with only marginal small oscillation (bottom right).}
   \label{fig3_s}
\end{figure}

The details of the horn crossing in phase space are illustrated in \Fig{fig3_s} where we plot the evolution of a localized semiclassical cloud initiated at the followed maximum before the crossing. The pertinent Poincare sections before and during the transition are plotted at the {\em cloud's energy} and the participating SPs are marked accordingly. It is evident that the horn crossing corresponds to a smooth diabatic swap from the pre-crossing maximum to the post-crossing saddle point that constitute the desired SP to follow. Thus, the only outcome of the avoided crossing  is a small oscillation about the new SP. This oscillation is a negligible effect compared to the stochastic spreading during the chaotic interval (see e.g. Fig.~4 of the manuscript). In terms of the population dynamics it corresponds to the barely noticeable oscillation taking place between the horn crossings in \Fig{fig1_s}b.

%%%%%%%%%%%%%%%%%%%%%%%%%%%%

%%%%%%%%%%%%%%%%%%%%%%%%%%%%
\section{Comment on the Poincare sections coordinates}
Note that the coordinate system of the Poincare sections is a Lambert conical projection of a sphere, where the radii are the meridians and the azimuth is the longitude Thus the origin corresponds to the north pole ($P_1=1$) and the $z=1$ circumference corresponds to the south pole (i.e. to the single state with $P_3=1$).

\clearpage
%%%%%%%%%%%%%%%%%%%%%%%%%%%%
\section{The Bogoliubov Frequencies}

The Bogoliubov procedure brings the Hamiltonian in the vicinity of the SP to the diagonalized form  
\beq
H \ \ \approx \ \ E[\text{SP}] + \sum_q \omega_q c_q^{\dag} c_q 
\eeq
The higher order non-linear terms are neglected.
The number of Bogoliubov Frequencies is the same as the number of degrees of freedom.
Due to conversation of the total particle number one frequency is zero (``zero mode").

From technical point of view it is convenient, for the purpose of finding the $\omega_q$ to write the 
equation of motion that is derived from the Hamiltonian as  ${\dot{z} = \mathbb{J} \partial H}$.   
Here $z$ is a set of canonical coordinates, $\partial$ denotes the derivatives with respect to those coordinates, and $\mathbb{J}$ is the symplectic matrix. For one degree of freedom ${z=(a,\bar{a})}$ and $\mathbb{J}$ 
is the second Pauli matrix. The generalization to more degrees of freedoms follows trivially.
The linearized equation of motion at the SP takes the form $\dot{z} = \mathbb{J} \mathcal{A} z$, 
where $\mathcal{A}$ is the Hessian, i.e. the matrix that consists of second derivatives with respect 
to the canonical coordinates.
The associated  characteristic equation is 
\beq \label{eA}
\det(\lambda - \mathbb{J} \mathcal{A}) \ \ = \ \ 0
\eeq
The eigenvalues are $\lambda_q= \pm i\omega_q$, 
where the $\pm$ distinguishes the frequency that refer to~$c_q$ 
from that of its conjugate $\bar{c}_q$. 
Eq(3) of the manuscript is the traditional version of presenting \Eq{eA} above.

%%%%%%%%%%%%%%%%%%%%%%%%%%%%
\section{Approximate Analytical Solution}

The characteristic polynomial \Eq{eA} is third order in $\omega^2$ implying the solutions come in three $\pm\omega_j$ pairs. Dropping the $\omega_0^2=0$ zero mode we are left with two pairs $\pm\omega_{1,2}$, corresponding to the two characteristic frequencies of the two-degree-of-freedoms system.
The latter are found by solving a quadratic polynomial in $\omega^2$. 
To get an analytical approximation, we estimate the followed SP population 
distribution ${\cal P}(SP)$ by the dark state solution of the linear problem, namely, 
\beq
P_1=\cos^2(\vartheta),~~P_2=0,~~P_3=\sin^2(\vartheta)
\eeq
where $\vartheta(x)=\arctan(\kappa_p(x)/\kappa_s(x))$ is the mixing angle.
The resulting solutions are,
\begin{equation}
\omega \ \ = \ \ 
\pm\left[\frac{b\mp \sqrt{D}}{2}\right]^{1/2}
\sqrt{\kappa_p^2+\kappa_s^2} 
\end{equation}
where,
\begin{eqnarray}
b(x) \ \ = \ \ (\varepsilon_e-\mu_e)^2+2\mu_e(\mu_e-2u_{\rm eff})
+\frac{3}{2}u^2_{\rm eff}(1+{\rm cos}^2(2\vartheta))+\frac{1}{2}, 
\end{eqnarray}
and the discriminant $D$ is,
\begin{eqnarray}
D(x) &=&  
9u_{{\rm eff}}^4{\rm cos}^2 (2\vartheta) -24\mu_{e}u^3_{{\rm eff}} {\rm cos}^2 (2\vartheta)
+u^2_{{\rm eff}}\Big[{\rm cos}^2 (2\vartheta)(16\mu^2_{e}-3)-3(\varepsilon_{e}-\mu_{e})^2(1+{\rm cos}^2 (2\vartheta))\Big] 
\nonumber \\
&& +u_{{\rm eff}}\Big[(\varepsilon_{e}+\mu_{e})+8\mu_{e}(\varepsilon_{e}-\mu_{e}) ^2-{\rm cos}(4\vartheta) (\varepsilon_{e}-3\mu_{e})\Big]
\nonumber \\
&& +\Big[ (\varepsilon_{e}-2\mu_{e})^2+(\varepsilon^2_{e}+\mu^2_{e})^2-4\varepsilon_{e}\mu_{e}(\varepsilon_{e}+\mu_{e})(\varepsilon_{e}-\mu_{e})-\mu^2_{e}(1+4\mu^2_{e})\Big]
\end{eqnarray}
Here
\beq
\varepsilon_{e} &=& \frac{\varepsilon}{\sqrt{\kappa_s^2(x)+\kappa^2_p(x)}} \\
\mu_{e} &=& \frac{\mu}{\sqrt{\kappa^2_s(x)+\kappa^2_p(x)}}
\eeq
are effective detuning and chemical potential, respectively. Comparison of these solutions with exact numerical diagonalization (see Fig.~3 of the manuscript) shows excellent agreement.

%%%%%%%%%%%%%%%%%%%%%%%%%%%%
\begin{figure}[h!] %  figure placement: here, top, bottom, or page
\centering
\includegraphics[width=10cm]{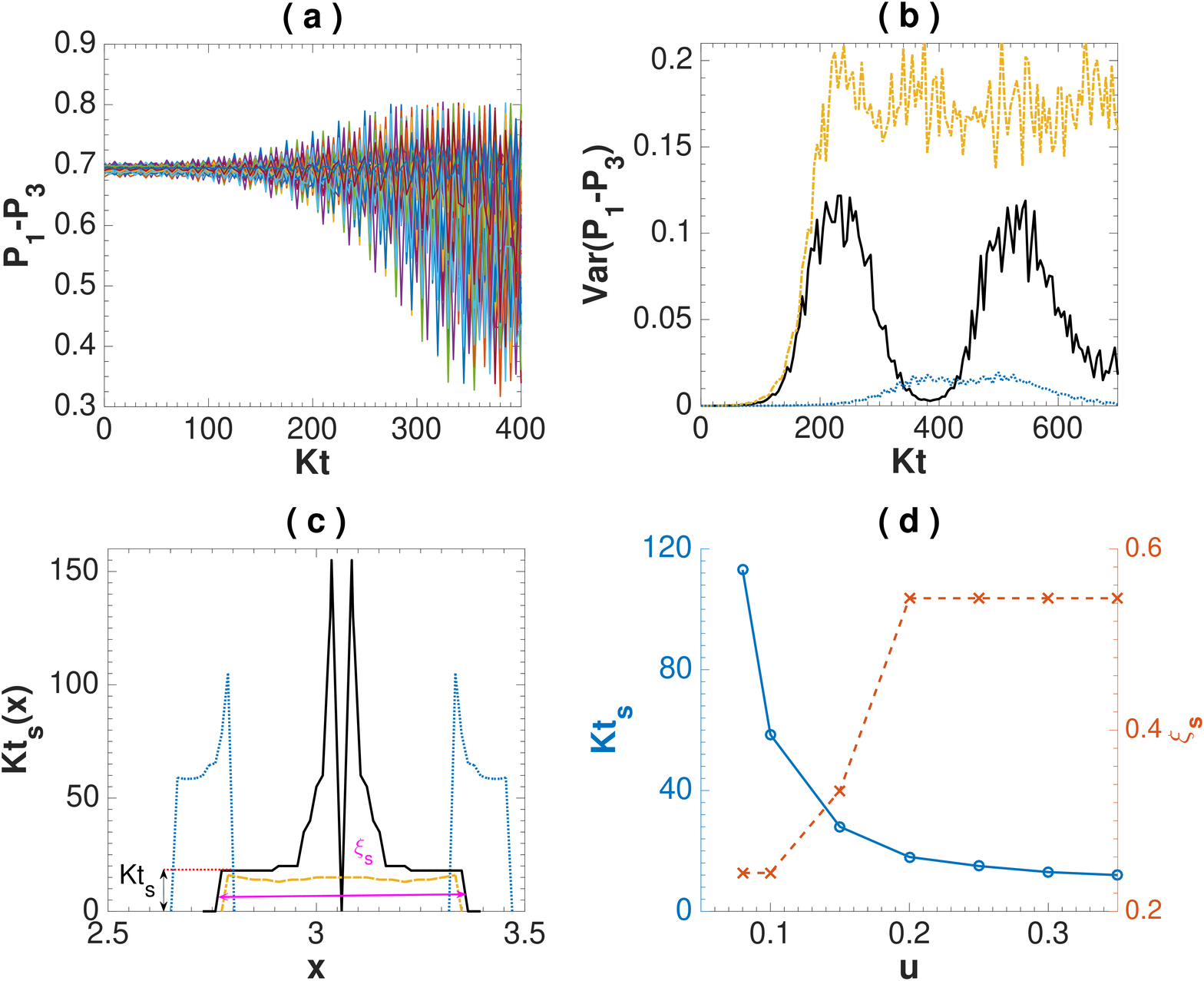} 
\caption{ (color online) Quasi-stationary spreading dynamics in
the chaotic strip. (a) Time evolution of the population imbalance
$P_1-P_3$ of representative trajectories in the semiclassical
cloud, for u=0.1 and fixed $x = 2.7273$. (b) The resulting variance
growth of the initially localized cloud for $u = 0.1$ (dotted
blue), 0.2 (solid black), 0.3 (dash-dotted orange), with fixed x within
the chaotic range. The spreading time $t_s$ to double the initial
variance is numerically extracted. (c) The dependence of the
spreading time $t_s$ on x for $u = 0.1, 0.2, 0.3$ (line types same
as in b). Outside of the chaotic intervals $t_s = 0$ means `no
spreading'. From here we find, for the given u, the chaotic
interval width $\xi_s$. The associated limiting $t_s$ is the minimal
value within this chaotic interval and marked in dotted red line. (d) The dependence of $\xi_s$
(x) and ts (o) on the interaction u.}
\label{fig4_s}
\end{figure}

%%%%%%%%%%%%%%%%%%%%%%%%%%%%
\section{Quasi-static spreading of a semiclassical cloud}

The procedure for obtaining the markers in Fig.~3c is illustrated in \Fig{fig4_s}. A semiclassical cloud, initiated around the followed SP, is propagated in time, with the parameters frozen at their instantaneous value at $x$. During the chaotic intervals the classical trajectories spread out, resulting in an increase of the population imbalance variance. The spreading is ballistic with a stochastic component.  At each $x$, we determine the time $t_s(x)$ at which the variance is doubled and $t_s$ is set to be the minimum of this characteristic spreading time during the chaotic interval. The duration of the chaotic interval $\xi_s$ is found from the region at which spreading takes place.  As shown in Fig.3, there is excellent agreement between these numerical results and the stability analysis prediction.

%%%%%%%%%%%%%%%%%%%%%%%%%%%%%%%%%%%%%%%%%%%%%%%%%%%%%%%%%%%%%%%%%%%%%%%%%%%%%%%%%%%%%%%%%%%%%%%%%%%%%%%%%%%
%%%%%%%%%%%%%%%%%%%%%%%%%%%%%%%%%%%%%%%%%%%%%%%%%%%%%%%%%%%%%%%%%%%%%%%%%%%%%%%%%%%%%%%%%%%%%%%%%%%%%%%%%%%
\clearpage
\end{document}